\documentclass[final,3p,times,twocolumn]{elsarticle}

\usepackage{amssymb}
\newcommand{\equref}[1]{{}{(\ref{#1})}}
\newcommand{\figref}[1]{{Figure~}{\ref{#1}}}
\newcommand{\tabref}[1]{{Table~}{\ref{#1}}}

\usepackage{amssymb}
\usepackage{multirow}
\usepackage{graphicx}
\usepackage{textcomp}
\usepackage{comment}
\usepackage{url}
\usepackage{gensymb}
\usepackage{placeins}
\usepackage{bm}
\usepackage{listings}
\biboptions{numbers,sort&compress}
 \usepackage{lineno}
\newenvironment{myfont}{\fontfamily{pcr}\selectfont}{\par}
\DeclareTextFontCommand{\textprogfont}{\myfont}

\journal{Applied Energy}

\begin{document}

\begin{frontmatter}



\title{Low-Cost Energy Meter Calibration Method for Measurement and Verification}


\author[eece]{Herman Carstens\corref{cor1}} 
\ead{hermancarstens@gmail.com}
\author[eece]{Xiaohua Xia} 
\author[ie]{Sarma Yadavalli}
\address[eece]{Centre for New Energy Systems (CNES), Department of Electrical, Electronic, and Computer Engineering, University of Pretoria, South Africa}
\address[ie]{Department of Industrial and Systems Engineering, University of Pretoria, South Africa}

\address{}

\begin{abstract}
Energy meters need to be calibrated for use in Measurement and Verification (M\&V) projects. However, calibration can be prohibitively expensive and affect project feasibility negatively. This study presents a novel low-cost in-situ meter data calibration technique using a relatively low accuracy commercial energy meter as a calibrator. Calibration is achieved by combining two machine learning tools: the SIMulation EXtrapolation (SIMEX) Measurement Error Model and Bayesian regression. The model is trained or calibrated on half-hourly building energy data for 24 hours. Measurements are then compared to the true values over the following months to verify the method. Results show that the hybrid method significantly improves parameter estimates and goodness of fit when compared to Ordinary Least Squares regression or standard SIMEX. This study also addresses the effect of mismeasurement in energy monitoring, and implements a powerful technique for mitigating the bias that arises because of it. Meters calibrated by the technique presented have adequate accuracy for most M\&V applications, at a significantly lower cost.

\end{abstract}


\end{frontmatter}

\section{Introduction}

Measurement and Verification (M\&V) is the process by which the savings from energy projects are independently quantified in a complete, conservative, consistent, transparent, and relevant manner~\cite{IPMVP}. M\&V is usually mandatory if projects are to be eligible for incentives such as credits or rebates. In many cases, limits are placed on the uncertainty with which savings can be reported~\cite{cdm_booklet, Ashrae2014, schiller2011national}. Following the International Standards Organization's Guide to the Expression of Uncertainty in Measurement (GUM)~\cite{isoguide, birch2003measurement} this uncertainty is usually expressed as a relative precision at a given statistical confidence level.

The challenging aspect of M\&V is that savings cannot be measured directly. Rather, a mathematical model of the energy systems' behaviour is created from measurements done prior to the intervention. This model may use covariates such as outside air temperature, occupancy, or production to characterise a facility's energy use. The  model then predicts what the energy use \textit{would have been} in the post-intervention period, had no intervention taken place. The difference between this predicted value and the actual measured energy use is the savings. 

\subsection{Definitions}

Various technical and closely related terms are used in this paper. Before proceeding, their definitions are clarified. \textit{Error} is the difference between the actual and the measured value. \textit{Random} errors are distributed symmetrically around the mean, and usually follow a normal distribution. \textit{Systemic} or non-random errors introduce bias. \textit{Bias} ``deprives a statistical result of representativeness by systematically distorting it"~\cite{dodge2010oxford}. For example, biased data will consistently have a different mean to the true mean. Random errors usually do not have this effect, except in the case of \textit{attenuation bias}, which will be discussed in Section~\ref{uncertainty_in_MV}.

\textit{Uncertainty} is ``the range or interval of doubt surrounding a measured or calculated value within which the true value is expected to fall with some degree of confidence"~\cite{Ashrae2014}.

\textit{Precision} relates to the ``fineness of discrimination"~\cite{birch2003measurement} or ``the closeness of agreement among repeated measurements of the same physical quantity"~\cite{Ashrae2014}. It is the uncertainty interval around a measured value, and should always be expressed with an associated statistical confidence. \textit{Confidence} is a probability, whereas precision is a distance, or size of the error band. Confidence and precision together usually define the broader term accuracy, which is ``the capability of an instrument to indicate the true value of a measured quantity"~\cite{Ashrae2014}. Note that the above definition of confidence, is popular although not technically correct~\cite{neyman1937outline, Ashrae2014, kacker2003use, kruschke2015doing} unless Bayesian methods are used.

By \textit{calibration} we mean the process of comparing an instrument to a standard or reference (instrument) to characterise its errors and improve its accuracy. The range and kinds of values that should be compared are often codified in standards. \textit{Disciplining} an instrument is a less complete calibration process where one only considers ranges and values expected to be encountered in a specific environment, and not the full range at which the instrument may be able to measure. Calibration is different from \textit{qualification}, which ensures the quality of an instrument model range, because of its design and manufacturing process. For example, tests are done to ensure the stability of meter readings under different environmental conditions, specified by the IEC~\cite{iec62053_21, iec62053_22, iec62053_23, iec60044_8}. Although a specific meter may be qualified because it is part of a model range and never lose this qualification, it may drift out of calibration.

\subsection{Uncertainty in M\&V} \label{uncertainty_in_MV}
During the M\&V process, three forms of uncertainty arise: measurement uncertainty, sampling uncertainty, and modelling uncertainty~\cite{IPMVP, Ashrae2014}. These will be addressed in turn.

\textit{Measurement uncertainty} refers to the difference between the actual and the measured values for a variable such as occupancy, outside air temperature, or energy. For projects where the interventions are spread over a large number of facilities, such as the residential mass rollout of energy efficient luminaires, it is not feasible to measure every home, and only a representative subset or sample is considered. This \textit{sampling uncertainty} needs to be quantified~\cite{ye2014optimal, carstens2014improvements, ye2016optimal}. \textit{Modelling uncertainty} arises because mathematical models to not reflect reality perfectly~\cite{wilkinson2010large, coakley2014review, kennedy2001bayesian}. Although some literature on sampling and modelling uncertainty exists~\cite{carstens2014improvements, ye2016optimal, olinga_cost} and a mathematical framework for M\&V has been constructed~\cite{xia2013mathematical}, measurement uncertainty is often neglected. For example, the The American Society of Heating, Refrigeration, and Air-Conditioning Engineers' (ASHRAE) Guideline 14 on Measurement of Energy, Demand, and Water Savings~\cite{Ashrae2014} assumes that data collected from US or Canadian  National weather services are measured without error~\cite{Ashrae2014}. This may be true for the immediate vicinity of the weather station, but not necessarily for the facility at which M\&V is done~\cite{sun2014uncertainty}. M\&V measurement instruments include surveys, questionnaires, inspection reports, and various kinds of meters. In this study, we will focus on metering uncertainty and calibration, and propose a method for keeping this uncertainty within acceptable bounds, at low cost.

The ASHRAE Guideline~\cite{Ashrae2014} combines the three kinds of uncertainties into a single figure, and does give uncertainty values for common instruments. However, this guideline assumes normally- or t-distributed parameter estimates and does not consider the errors-in-variables effect, on which we will elaborate below. Other leading guidelines mention measurement error, but do not discuss its more detrimental effects~\cite{calimv, see_action_guide, bonneville2012regression}. A notable exception is the Uniform Methods Project~\cite{umpch13, umpch23}, chapters 13 and 23. The Clean Development Mechanism (CDM) guidelines also use knock-down factors to account for  measurement uncertainty~\cite{cdmeb73a04}.

It has been shown that assuming that measurement error is negligible is valid for cases where metering is done on a sample of a population with normal to high variance~\cite{carstens2015measurement}. However, in cases where sampling uncertainty does not dominate measurement uncertainty, for example for single-facility studies or where all facilities are metered, the uncertainty in the meter data becomes significant in the overall uncertainty calculation. In such cases, measurement uncertainty may make a material difference to overall reporting uncertainty. Yet in all cases the reduction of measurement uncertainty through meter calibration is costly, not only because of laboratory fees, but also because of meter installation and removal costs.

A study of the present state of the art regarding measurement uncertainty in energy monitoring has been conducted~\cite{carstens2016measurement2}, although it has not yet been published at the time of writing. One of the key findings relevant to this research is that the little-known errors-in-variables effect may be significant in some M\&V cases. Briefly, conventional thinking is that bias in the measurements will bias the model, while zero-mean noise in the measurements will not bias the model. However, when unbiased noise in the measurement of the independent variables is present, it leads to biased (``attenuated") parameter estimates when these data are used for modelling~\cite{ridge1997errors, umpch13, umpch23, sonnenblick1995framework}. This is the errors-in-variables effect. There are various methods of reducing this bias~\cite{carroll2006measurement, gustafson2003measurement, fuller2006measurement}, and some of them will be implemented below.

\subsection{Calibration in M\&V}
One way to circumvent or mitigate measurement uncertainty is to use accurate, calibrated meters. One then assumes that the measurement uncertainty is negligible. This is the approach taken by South Africa's 12L tax incentive programme~\cite{twelveL}, where meters are required to be calibrated by an accredited laboratory at fixed intervals. Other international programmes adopt similar approaches~\cite{ahmad2016building}. This is a sound principle from a regulatory point of view. It minimises the consumer's risk, that is, the risk of using an inaccurate meter. However, a significant opportunity cost is incurred because many projects are never implemented due to monitoring, laboratory, and plant shut-down costs. An example of this has been recorded for the CDM lighting retrofit project specifications~\cite{shishlov2014review, michaelowa2009challenges}. Striking a balance between calibration costs and monitoring accuracy is, therefore, an important but non-trivial consideration for policy makers.

Our method also addresses a second calibration difficulty. The European Measurement Instrument Directive~(MID)~\cite{ec_directive} requires that meters be calibrated in-situ, that is, in the environment in which they will be installed~\cite{femine2009advanced}. Besides regulatory compliance in European countries, a method capable of doing this is also convenient and practical. Various solutions have been proposed, from travelling laboratory-grade instruments with metrologists~\cite{femine2009advanced} to add-on calibrators~\cite{amicone2009smart}. However, these solutions entail high costs and specialised equipment. Because in-situ ``calibration" does not test at all meter levels, but only at those experienced during the measurement period, we will sometimes refer to our method as ``disciplining" or ``verifying" the Unit Under Test (UUT)~\cite{somppi2007case}. However, in mismeasurement statistics, the term ``calibration" is often used to describe the procedure of correcting mismeasured data. For example, one method similar to the one proposed in this paper is called ``Regression Calibration"~\cite{carroll2006measurement}.

Commercial calibration techniques usually rely on having calibrators that are at least four times as precise as the UUT. This is called the Test Uncertainty Ratio (TUR)~\cite{somppi2007case}. Others focus on accept/reject decisions~\cite{deaver1993maintain}. The other low-cost calibration option is to use a PC and Data Acquisition (DAQ) board-based system. It has been demonstrated that such systems can achieve impressive accuracies at a fraction of the cost of commercial standards~\cite{cataliotti2011uncertainty, cataliotti2015daq}, in a research environment. DAQ-based calibrators are set to become popular in future, although the technology probably needs more time to become commercialised. 

One of the reasons imprecise reference instruments are avoided is because it will lead to an error-in-variables effect, requiring Measurement Error Models (MEMs). To the best of our knowledge, MEMs have not been applied to electrical meter calibration before. We will also use the Bayesian approach below. Although Bayesian approaches can be applied to certain mismeasurement problems~\cite{gustafson2003measurement} and are becoming popular in M\&V~\cite{walter2014uncertainty, shonder2012bayesian, claridge2014methodologies} and metrology generally~\cite{lira2016gum, rossi2003probabilistic, cox2008probabilistic, riddle2014guide}, the way in which we apply it may also be novel. A second reason that imprecise reference instruments may be used for the problem under investigation is that measurement uncertainty for M\&V is often dominated by other forms of uncertainty such as sampling, as mentioned before. The goal of disciplining the meter for such cases is different from that of a calibration laboratory calibrating a meter; it is simply to keep measurement uncertainty as a negligible component of overall uncertainty.

The method proposed in this paper is therefore novel for a number of reasons. Calibration is usually done in a laboratory, using highly accurate and expensive laboratory equipment, whereas this method will use a commercial-grade meter as a calibrator. Calibration usually does not account for errors in the calibrator, whereas this method will do so. To our knowledge, Simulation Extrapolation has not been used for meter calibration, and has also not been combined with Bayesian regression as is done in this paper. Finally, the proposed approach provides a more practical solution to in-situ calibration than those proposed in literature.

This paper is structured as follows. Section~\ref{lowcostcal} investigates a low-cost calibration (disciplining) technique. Error classification is discussed and applied to the kinds of errors found in energy meters. An MEM is then selected. Section~\ref{SIMEX_application} applies this MEM to actual data and evaluates its effectiveness in parameter estimation. Section~\ref{application} broadens the scope of the calibration context and makes refinements using the Bayesian approach. Finally, the results are discussed and we draw conclusions.

\section{Developing a Low-Cost Meter Calibration Algorithm} \label{lowcostcal}
Given that meters need to be verified but that this can be prohibitively expensive, the possibility of disciplining an installed meter with another commercial (rather than laboratory) accuracy meter should be investigated. 

The cost saving from using the method proposed in this paper will vary with the number of meters disciplined instead of being sent to a calibration laboratory. The cost saving for the client will also vary with the cost of facility down-time needed to install and remove meters. The meters needed when using the proposed method are not more or less accurate than standard energy meters, and their accuracy will normally be determined by other factors than the method proposed.

The commercial meter-as-calibrator will measure with a non-negligible error. A range of scenario-specific MEMs has been developed to account for the ways in which the measurement errors may arise. The nature of the errors needs to be classified accurately to apply the correct MEM to a problem. In some cases, certain simplifying assumptions may restrict the model's applicability. In others, incorrect assumptions may lead to erroneous results. Mismeasurement in M\&V is treated more fully in previous work~\cite{carstens2016measurement2}, and Carroll et al.~\cite{carroll2006measurement} and Gustafson~\cite{gustafson2003measurement} have written excellent textbooks on the problem.
%National requirements such as the South African 12L tax incentive~\cite{twelveL} require that meters used for M\&V projects eligible for such programmes be calibrated by a laboratory. This is a sound principle, but the cost of calibration is disproportionate to the uncertainty contribution made by meters~\cite{carstens2015measurement}. Therefore low cost calibration of meters represents an M\&V cost reduction opportunity.

We will use $\mathbf{x}$ to denote the true values of the independent variable (reference instrument or calibrator) and $\mathbf{y}$ the true values of the dependent variable (UUT). To differentiate between the true values and the observed values which are measured with error, we use an asterisk (*) for measured values. Since $p$ is often used to denote precision, and $P$ to denote power, we use $\pi$ to denote probability.

Before looking at the errors themselves, two related concepts need to be mentioned. An \textit{exposure model} is often needed when specifying an~MEM. Although we often have a model of how errors arise in the form $f(\mathbf{x}^*|\mathbf{x})$, we cannot work backwards to infer $\mathbf{x}$ from the observed $\mathbf{x}^*$. An exposure model describes this function: $f(\mathbf{x}|\mathbf{x}^*)$. This is often done through a third variable $\mathbf{z}$. The exposure model then takes the form $f(\mathbf{x}|\mathbf{z}$), where $\mathbf{z}$ is some covariate measured without error. %Exposure models may be used in energy monitoring when we may measure how temperature affects our instrument, but we may not have exact past temperature measurements for our instrument when the data were collected.
%A possible application of exposure models in energy monitoring would be the exposure of subjects to awareness campaigns.

\textit{Model identifiability} is another concern. Sometimes a key piece of information is missing, and the data are not enough to identify all the model parameters uniquely. Carrol et al.~\cite{carroll2006measurement} and Gustafson~\cite{gustafson2003measurement} adopt complementary approaches. Briefly, Gustafson found that non-identifiability is not always detrimental, and Carroll et al. found that identifiability is not always good enough, especially for threshold cases. Gustafson also found that specifying uncertainty (priors) on some parameters may even lead to better results than fixing those parameters at slightly incorrect values for the sake of identifiability.

\subsection{Error Taxonomy}

Errors may vary in a number of ways. First, errors can be \textbf{correlated or uncorrelated}. This is not in the same category as the classifications that follow but is an important distinction nonetheless. Errors that are uncorrelated with other variables are the simplest to model. Consecutive errors may also be autocorrelated in a time series. This sequentiality is hidden in scatter plots and regression analysis, although it still affects the estimates.

Errors can be \textbf{classical or Berkson}. Classical errors take the form $\mathbf{x}^{*}=\mathbf{x}+\bm\epsilon$, and are more common. This is when the error is in the instrument itself. Berkson errors take the form $\mathbf{x}=\mathbf x^*+\bm\epsilon$. This occurs when the actual value of the measurand varies around the assigned, or measured value because the source of the error is external to the instrument.

Errors are classified as \textbf{multiplicative or additive}. Multiplicative errors are of the form $\mathbf{x}^{*} = \mathbf{x}\bm{\epsilon}$, whereas additive errors take the form $\mathbf{x}^{*} = \mathbf{x}+\bm{\epsilon}$. The additive error assumption is a popular one as it greatly simplifies MEM mathematics: additive errors are usually associated with constant variance throughout the measurand range. This is called \textit{homoscedasticity} and is a critical assumption when performing Linear Regression (LR). The majority of techniques have been developed to deal with this kind of model. However, this assumption is not always valid. For example, it has been demonstrated that energy meter measurement errors are non-linear and multiplicative~\cite{carobbi2009error}, and are thus \textit{heteroscedastic}. This has been acknowledged to produce problems in econometric energy analyses~\cite{schiller2011national}, and frequentist methods to account for some cases in regression analysis has been developed~\cite{carroll1988transformation}. It may be mitigated by assuming a lognormal distribution and working with $log\mathbf{x}^{*}$, since $log\mathbf{x}\bm{\epsilon}=log\mathbf{x}+log\bm{\epsilon}$, transforming the error model to an additive one. However, the assumption of a lognormal distribution on $\mathbf{\epsilon}$ (so that $log\epsilon \sim$ Normal), although mathematically convenient, is not always valid or preferred~\cite{carroll2006measurement}. Heteroscedasticity can be present even for additive errors when they have non-constant bounds over the measurement range, such as energy meters and Current Transformers (CTs)~\cite{iec62053_21, iec62053_22, iec62053_23, iec60044_8}. These bounds are shown in \tabref{tab:Meterspec}.
 
Errors may be \textbf{differential or non-differential}. Non-differential errors mean that $\mathbf{x}^{*}$ contains no more information about $\mathbf{y}$ than $\mathbf{x}$ does. The response does not change due to measurement. Differential errors may occur when the response~$\mathbf{y}$~is measured before the covariates~$\mathbf{x}^{*}$~and~$\mathbf{z}$, and these variables are liable to change. For example, the diet ($\mathbf{x}$) of women with breast cancer may be measured only after their diagnosis $\mathbf{y}$. It is possible that the test subjects change their diet as a result of the diagnosis~\cite{carroll2006measurement}. Another example is when $\mathbf{x}^*$ is a proxy for $\mathbf{x}$, not simply a mismeasurement. For example, plug loads are sometimes used as a proxy for occupancy~\cite{ward2016exploring}. Differential errors may also occur in ex-post energy use surveys for residential retrofit programmes where the response (purchasing of certain equipment, for example) is measured before other variables of interest are measured.

Last, the function $\mathbf{y}(\mathbf{x})$ may be \textbf{linear or nonlinear}. This is not an assumption about the errors themselves but does affect the kinds of errors that are permissible. The linear assumption is popular as it allows LR to be used if one assumes normally distributed additive errors. For many models, this is a valid assumption. However, Carobbi, Pellicci, and Vieri~\cite{carobbi2009error} have shown that the standard $P=VI$ electrical power equation, where $P$ is Power in Watts, $V$ is potential difference in Volts, and $I$ is current in Amperes, can be modelled as
\begin{equation} \label{equ:carobbi_error}
P_{n}=(1+\alpha)VIcos(\phi+\phi_{c})+\epsilon,
\end{equation}
when an energy meter measures with error. In this equation, $\alpha$ is the gain error, $\phi_c$ is the phase error, and $\epsilon$ is the bias error. The gain error $\alpha$ changes the amplitude of measured power fluctuations, but does not affect the mean. In other words, the larger the energy reading, the larger the error. The size of this error is directly proportional to the magnitude of the energy reading. The bias error $\epsilon$ offsets the measured power, changing the mean power read by the meter, but does not change the amplitude of the fluctuations. This error may bias the power and energy reading upwards or downwards. The phase error $\phi_c$ has a similar net effect to the gain error, but changes according to the power factor error of the meter. The power factor is the ratio of real to apparent power. At a unity power factor, the real power in Watts is equal to the apparent power in Volt-Amperes, so that the $P=VI$ equation holds: power in Watts truly is equal to Volts multiplied by Amperes. However, as the current and potential difference move out of phase, the power factor changes, as this changes the real-to-reactive power ratio. This phase difference is expressed in radians. Non-unity power factors are very common, and are caused by electrical motors and power electronic circuits, which usually have inductive loads. Mismeasuring the power factor will have the net effect of changing the gain of the meter. Carobbi, Pellici, and Vieri's contribution~\cite{carobbi2009error} was to show that~\equref{equ:carobbi_error} is a statistically adequate model, capturing the real error behaviour of energy meters without specifying too many parameters. 

Although this error is multiplicative, the error bounds in the IEC meter qualification standards~\cite{iec62053_21, iec62053_22, iec62053_23} are additive. The meter may still have a multiplicative error, but this error is always smaller than the additive error bound. In cases where these are the only data available, additive errors may have to be assumed. Furthermore, the error model is only non-linear if the phase error term $\phi_c$ is of interest.

\subsection{Meter Calibration}

The method below focusses on energy meters, but can be used for instruments measuring other parameters as well. The most analogous cases are flow measurement~\cite{crenna2009probabilistic}, and possibly exhaust gas analysis~\cite{pendrill2006exhaust}. Occupancy measurement may also benefit from thoughtful application~\cite{ward2016exploring, wang2016meta}, but temperature measurements are often biased due to spatial variations~\cite{sun2014urban}, and will require more careful application.

The proposed approach is to discipline a meter (the UUT) using another relatively low-specification commercial-grade metering system. This could be done by installing the meters in parallel in-situ at the facility for a short period, such as 24 hours. The data from the calibrator are then used to correct (discipline or calibrate) the data from the UUT. Although the UUT is not calibrated, we assume that it is of reasonable quality. For example, the model range to which the UUT belongs should be qualified to an IEC specification. This is necessary to ensure that readings will remain stable under different operating conditions such as winter and summer temperatures.

For high-accuracy laboratory multimeters measuring to six or eight decimal places, various additional factors should be considered during calibration. These include thermoelectric voltages, cable impedance, and performance at different frequencies~\cite{ukasm3003}. However, these fluctuations are small enough to be negligible for commercial energy measurement applications.

When an imprecise reference is used to quantify an imprecise UUT, a Measurement~Error~Model~(MEM) or `errors-in-variables' model has to be used. For example, suppose that the output of a power supply is measured with a reference meter (\textbf{x}) and a Unit Under Test (UUT) (\textbf{y}). If both the reference and the UUT are perfectly accurate, a regression line with a gradient of one should be drawn on the \textit{xy} plane. If only the UUT has an error (thus an error in the response or dependent variable measurement), the dependent variable $\mathbf{y}^{*}=\mathbf{y}+\epsilon$ will be measured by the UUT. This kind of error will add noise (vertical scatter), but should not bias the result.

However, when the errors are in the independent or input variable from the reference (\textbf{x}), the effect is more insiduous. If \textbf{x} is measured with random, zero-mean error, the result is not increased scatter, but bias. This can be visualised by seeing the \textit{x}-axis ``spread out", flattening the slope of the regression line and biasing the y-intercept upwards. Consider the meter error equation~\equref{equ:carobbi_error}. If there is random error in~\textbf{x} (or $VI$ in this case), we expect the y-intercept ($P$-intercept in this case), to move upwards. The bias error $\epsilon$ will therefore be~\textit{overestimated}. Also, since the slope flattens, the gain error $\alpha$ is expected to be \textit{underestimated}. However, there is a complicating factor: this is not straight-line regression. There is a non-linear and confounding term in $cos(\phi + \phi_c)$. This illustrates that the effect of mismeasurement for more complex cases than straight line regression is that the parameter estimation bias for individual parameters is unpredictable. The gain error could be overestimated and the phase error underestimated. It is not possible to predict this beforehand, which is partly why mismeasurement is such a difficult problem to address. Two other effects compound the problem. The first is that with errors in $\mathbf{x}$, the standard errors on the parameter estimates become unreasonably small. This means than not only is the parameter estimate biased, but the apparent confidence interval around this biased value is too narrow. The third effect of random errors in \textbf{x} is that non-linear features become obscured~\cite{carroll2006measurement}. For example, a certain amount of vertical scatter in a sinusoidal graph will not hide its sinusoidal shape. However, the same amount of horizontal scatter will make the function appear as a horizontal cloud. This effect holds for all non-linear functions.

%For our case, our approach will be as follows. First, we will investigate the nature of the errors in \textbf{x} and \textbf{y}, then propose an MEM approach, compare different implementations of it, and solve for the parameters of interest. We will then use these parameter estimates to solve for a real-world scenario.

\subsection{Errors in $\mathbf{x}$}
\begin{table} 
\begin{center}
\caption{Accuracy specification for IEC Class 3 meter~\cite{iec62053_23}. $P_n$ denotes the rated power, $I_n$ rated current, and $I_{max}$ the maximum current.}
\label{tab:Meterspec}
\begin{tabular}{c c c}
\hline 
\textbf{Value of Current} & \textbf{Power Factor} & \textbf{Error limit} \\ 
\hline
$0.02 I_n \leq I \leq 0.05 I_n$ & 1 & $\pm 0.04P_n$ \\ 
$0.05 I_n \leq I \leq I_{max}$ & 1 & $\pm 0.03P_n$ \\ 
$0.05 I_n \leq I \leq 0.1 I_{n}$ & 0.5 & $\pm 0.04P_n$ \\ 
$0.1 I_n \leq I \leq I_{max}$ & 0.5 & $\pm 0.03P_n$ \\ 
\hline 
\end{tabular} 
\end{center}
\end{table}

The calibrator data is selected for the \textit{x}-axis, rather than the UUT. This is because the calibrator should have smaller errors than the UUT. In this way, attenuation bias is minimised as much as possible before adjustments are made.

To be conservative, we select the highest (least accurate) IEC class meter and Current Transformer (CT) combination for our reference instrument. This would be a Class 3 meter~\cite{iec62053_23} with a Class 5 CT~\cite{iec60044_8}. The meter accuracy limits are shown in \tabref{tab:Meterspec}. For power factors between $\pm 0.5$ and 1, the accuracy limits were linearly interpolated. The CT has a flat accuracy limit of 5\% of the rated current. We note that these are additive error bounds relative to the rated, or full scale, current. We assume that this meter is calibrated. The true errors may still be multiplicative, but will fall within these additive bounds. 
Metrology guidelines often recommend that a uniform error distribution between the error bounds be assumed~\cite{birch2003measurement}. However, this is too conservative. Instead, errors bounds are assumed to be the 95\% confidence limits on a normal distribution~\cite{birch2003measurement, Ashrae}.  The readings are also assumed to be unbiased. Errors are assumed to be classical, non-differential, and uncorrelated. Even though errors are additive, they are heteroscedastic (having non-constant variances) due to the stepwise nature of the error bounds as described by~\tabref{tab:Meterspec}. The total error would be the root sum of squares of the meter and CT error bounds at a given point:
\begin{equation} \label{equ:sigma_u}
p_{combined} = \sqrt{p_{meter}^2(x)+p_{CT}^2(x)}.
\end{equation}
Let $p_{combined}(x)$ be the combined error bound at $x$, and $z$ be the standard score (or coverage factor). The standard deviation on the a given reading can then be written as
\begin{equation} 
\sigma_u = \frac{p_{combined}(x)}{z}.
\end{equation}

The rated power of the meter is assumed to be 200~kW, and the rated current for the CT is assumed to correspond to this value.

The measured values on the calibrator $\mathbf{x}^*$ can then be defined as

\begin{equation}\label{equ:x_star_given_x}
\mathbf{x}^* \sim Normal(\mathbf{x}, \sigma_u)
\end{equation}

\subsection{Errors in $\mathbf{y}$}
For errors in our Unit Under Test ($\mathbf{y}$) we may make more detailed assumptions. Following Carobbi et al.~\cite{carobbi2009error}, we assume that the characteristic function for the UUT is
\begin{equation} \label{equ:carobbi_error_y_star}
\mathbf{y}^*=(1+\alpha)\mathbf{x}cos(\phi+\phi_{c})+\epsilon,
\end{equation}
where $\alpha$ is the gain error, $\phi$ is the phase difference between voltage and current, $\phi_c$ is the phase error, and $\epsilon$ is the bias error. The errors are classical, with multiplicative and additive components. They are also homoscedastic, and the function is non-linear. Since these errors will not cause attenuation bias, the MEM is not selected on their basis. However, they are built into the overall measurement model.

\subsection{MEM selection}
Since $\phi_c$ is one of the variables of interest, this is a non-linear function, and that standard LR techniques such as Fuller's method of moments~\cite{fuller2006measurement} are not valid unless the $cos(\phi+\phi_c)$ term in \equref{equ:carobbi_error} is neglected. 

Although $f(\mathbf{x}^* | \mathbf{x})$ is available by~\equref{equ:x_star_given_x} in the form of a distribution function, $f(\mathbf{x} | \mathbf{x}^*)$ is not. To obtain this, we would need an exposure model, which is not at our disposal.
One approach would be to specify a na\"ive Bayesian model on the data using~\equref{equ:x_star_given_x}. By specifying a distribution on $\mathbf{x^*}$, the noisy independent variable is taken into account, mitigating the attenuation effect to some degree. If errors were Berkson rather than classical, this would be accurate. However, this is not the case for our measurements.

Since we do not assume the availability of an exposure model, repeated measurements, or a sub-set of gold-standard measurements, MEMs like Regression Calibration, Maximum Likelihood techniques, and the Bayesian approach are not available to us. Instead, we propose a hybrid SIMulation EXtrapolation (SIMEX) solution.

\subsection{SIMEX} \label{SIMEX_section}
SIMEX is a simple, powerful algorithm that compensates for measurement error using only $f(\mathbf{x}^* | \mathbf{x})$ in the form of $\sigma_u$. It was first proposed by Cook and Stefanski~\cite{cook1994simulation}, and a useful summary can be found in Carroll, Stefanski, et al.~\cite{carroll2006measurement}. Since this method is easily automated, it can be classified as a machine learning algorithm. The premise is that although the biased parameter estimates $\{\alpha^*, \phi^*_c, \epsilon^*\} = \bm{\theta^*}|\mathbf{x}^*$ cannot be unbiased directly, they can be biased even more by adding more noise to $\mathbf{x}^*$. By repeating this biasing for increasing noise levels, the relationship between noise in $\mathbf{x}$ and bias in $\bm{\theta}$ is found. A trend can be observed from these successive noise levels, and the noise-free state $\bm{\theta} | \mathbf{x}$ can then be inferred by backwards extrapolation.  \figref{fig:simex_illustration} illustrates this graphically. The SIMEX procedure can be defined more rigorously as follows:

\begin{enumerate}
\item Describe the variance $\sigma_u$ due to mismeasurement. 
\item Describe the UUT function $\mathbf{y} = f(\mathbf{x})$. 
\item Specify the vector of noise multiples to obtain a vector $\bm{\zeta}$ of length $n$ at which simulation will be done. Values for $\bm\zeta$ can start at zero and could go up to five.
%\item Select a sample size $n$: the number of simulations that will be done at each noise level. Alternatively, select the number of points along $\bm{\zeta}$ at which parameter estimates are calculated.
\item Calculate $\mathbf{x}_{\bm{\zeta},\mathbf{n}}^* = \mathbf{x}^* + (1+ \sqrt{\bm{\zeta}})\sigma_u$. The reason for the square root on $\zeta$ is explained by Carroll et al.~\cite{carroll2006measurement}, but is beyond the scope of this study.
\item Solve $\mathbf{y}_{\bm{\zeta},\mathbf{n}}^* = f(\mathbf{x}_{\bm{\zeta},\mathbf{n}}^*)$ to find $\bm{\theta}(\bm\zeta)$. If $f(\mathbf{x})$ is linear, this can be done by LR. For non-linear problems, an appropriate function should be specified, and an optimisation algorithm is needed to solve for the function parameters.
\item \label{second_regression} For every element of $\bm\theta$ (that is, $\alpha, \phi_c, \epsilon$), a vector of $n$ solutions in $\bm{\zeta}$ is now available. Consider the gain error $\alpha$. If the function $\hat \alpha (\bm\zeta)$ were linear, one could now solve
\begin{equation} \label{equ:alpha_hat_ito_zeta}
\hat \alpha (\bm\zeta) = a_\alpha\bm\zeta + b_\alpha.
\end{equation}
Carroll et al.~\cite{carroll2006measurement} divided $\bm\zeta$ into discrete levels with many samples per level. They then used the mean of every level of $\bm\zeta$. However, since this is not an expensive step, one would rather regress against the full data set than assume that the distribution is symmetric. Also, rather than using discrete levels, we prefer a linear spacing of points between the maximum and minimum values of $\bm\zeta$.

\item \label{solving_for_alpha} The unbiased parameter estimate $\alpha|\mathbf{x}$ is found by solving \equref{equ:alpha_hat_ito_zeta} for $\zeta=-1$. This is illustrated graphically in \figref{fig:simex_illustration}.
\item Repeat Step \ref{solving_for_alpha} for $\phi$ and $\epsilon$.
\end{enumerate}

\section{Case Study: SIMEX Application} \label{SIMEX_application}

\begin{figure*}[tb]
\begin{center}
%\vspace{5cm}
\includegraphics[width=0.8\textwidth]{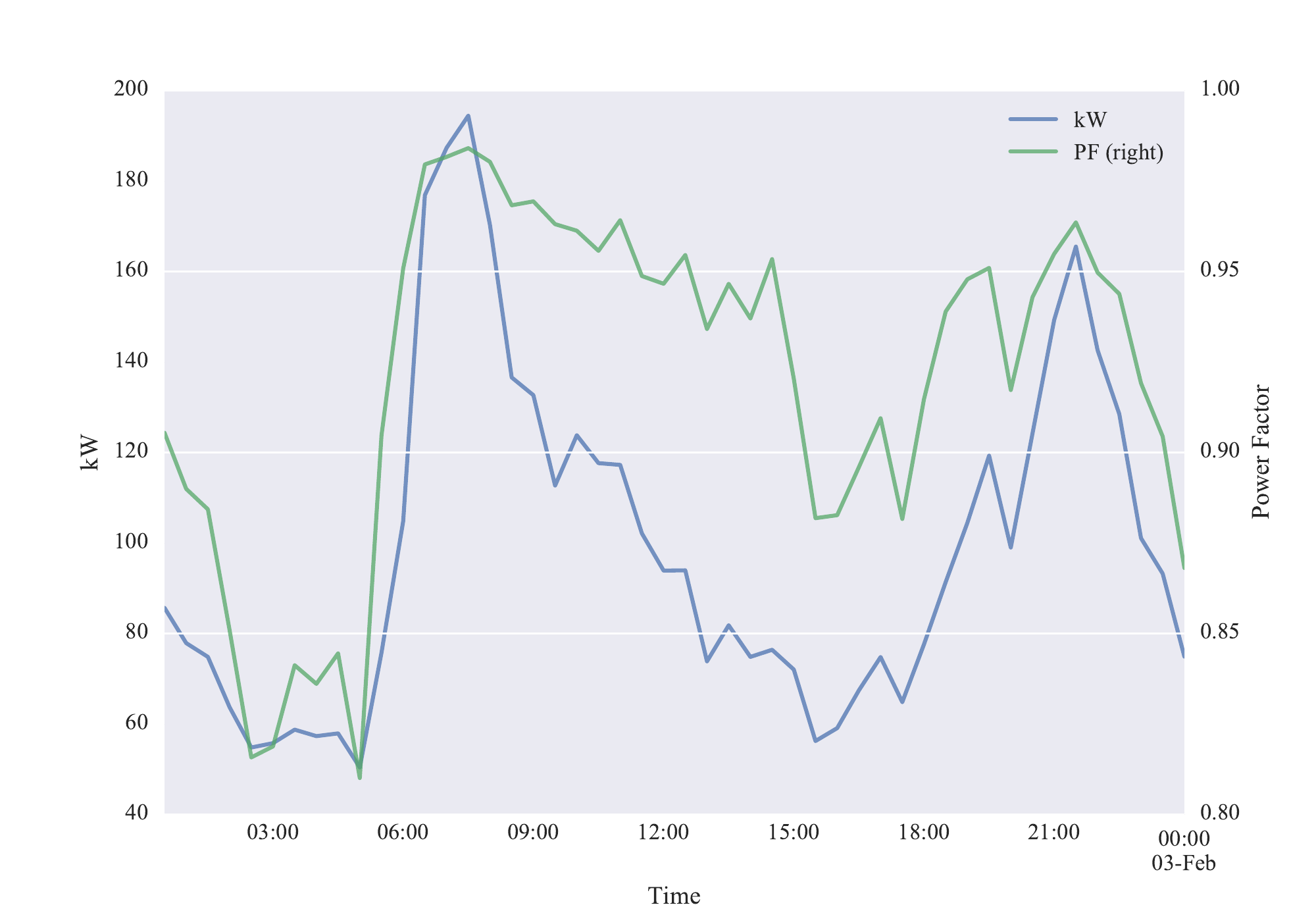}    % The printed column
\caption{Load (kW) and power factor (PF) profile for the period used for calibration.}  % width is 8.4 cm.
\label{fig:load_profile}                                 % Size the figures
\end{center}                                 % accordingly.
\end{figure*}

The SIMEX algorithm was modified slightly and applied to the meter calibration problem at hand. Initially, we tested the algorithm with an energy data set of linearly interpolated points between 0~and~$I_n$, at three different power factor levels. This simulates a  laboratory set-up. However, to simulate in-situ calibration, real load profile data was needed. We used the actual energy consumption of a university residence at the University of Pretoria, on 2 February 2016. The data are plotted in \figref{fig:load_profile}. The power factor was converted to a phase angle by $\theta = cos^{-1}(Power\ Factor)$.

One problem with such data is that power factor and energy use are correlated. High power factors occur at high loads, and low power factors occur at lower loads. This could be due to heavy loads such as geysers having unity power factors and forcing the overall power factor upwards during peak times. Such a correlation has a confounding effect on parameter estimation of $\phi$ especially. Using larger calibration data sets such as a one-week rather than a one-day period helps only marginally since the system still has the same correlation characteristics.

For our experiment, the (unknown) parameter values are set as shown in~\tabref{tab:params}, and altered the data using~\equref{equ:x_star_given_x} and~\equref{equ:carobbi_error_y_star} to produce the observed data $\mathbf{x}^*$ and $\mathbf{y}^*$.
\begin{table}[tb]
\centering
\caption{Parameter values}
\label{tab:params}
\begin{tabular}{c c c}
\hline
\textbf{Parameter name} & \textbf{Symbol} & \textbf{Value}\\
\hline
Gain Error &$\alpha$& 0.2\\
Phase Error&$\phi_{c}$& 0.2\\
Bias Error &$\epsilon$& $\sim$ Normal(5, 2.5)\\
\hline
\end{tabular}
\end{table}
The SIMEX algorithm was implemented in the following manner, according to the steps described in Section \ref{SIMEX_section}:
\begin{enumerate}
\item The variance $\sigma_u$ is described by~\equref{equ:sigma_u}.
\item The UUT function $\mathbf{y}^* = f(\mathbf{x})$ is described by~\equref{equ:carobbi_error_y_star}.
\item The SIMEX graphs were found to be non-linear, especially for $\zeta$ values above 2. Therefore, $n=300$ points between $\zeta=0.5$ and $\zeta=5$ were selected. Points between 0 and 0.5 were not included because in this region the data converge asymptotically to $\zeta=0$, which is an artefact of the algorithm rather than a real trend. 
\item These $n$ realisations were generated using Python's \textprogfont{numpy} library~\cite{van2011numpy} and the \textprogfont{numpy.random.normal} pseudo-random number generator for 
\begin{equation}
\mathbf{x}^*_{\bm{\zeta},\mathbf{n}} \sim Normal(\mathbf{x}^*,\sigma_u).
\end{equation}
The variance $\sigma_u$ was defined by~\equref{equ:x_star_given_x}.
\item In this case, we implemented Python's \textprogfont{scipy}~\cite{oliphantscipy} module to find the least-squares solution of \equref{equ:carobbi_error_y_star} for $\bm{\bm\theta}(\bm{\zeta})$. The library implements the Broyden et al. quasi-Newton method~\cite{nocedal2006numerical} by default. Non-default optimisation algorithms were also tried but showed poorer convergence and efficiency.
\item A non-linear model was assumed to solve for $\bm{\theta(\zeta)}$. The data exhibit a sigmoid shape, and various sigmoid-shaped functions such as piecewise linear, hyperbolic tangent, sinusoid, and logistic functions were tested. The standard logistic function below delivered the most reliable results. For $\alpha$, for example, one would solve
\begin{equation} \label{equ:alpha_logistic}
\hat \alpha (\bm\zeta) = \frac{L_\alpha}{1+e^{k_\alpha(\bm\zeta-\zeta_{0,\alpha})}}
\end{equation}
for $L_\alpha, k_\alpha, $ and $\zeta_{0, \alpha}$. $L$  determines the curve's maximum value, $k$ determines the slope, and $\zeta_0$ determines the $x$-value of the midpoint. The data and resultant fit for one realisation can be seen in \figref{fig:simex_illustration}. The same optimization algorithm as the previous step was used.
%The first is standard LR via the Python \textprogfont{scikit-learn} library~\cite{pedregosa2011scikit}. Because these estimates may be used in risk calculations later, a Bayesian solution was also implemented. Space does not permit a full explanation of the Bayesian approach. Several excellent guides have been written. Kruschke~\cite{kruschke2015doing} is most notable, although more applied Bayesian measurement tutorials are also worth reading~\cite{riddle2014guide, cox2008probabilistic}
\begin{figure*}[tb]
\begin{center}
%\vspace{5cm}
\includegraphics[width=0.8\textwidth]{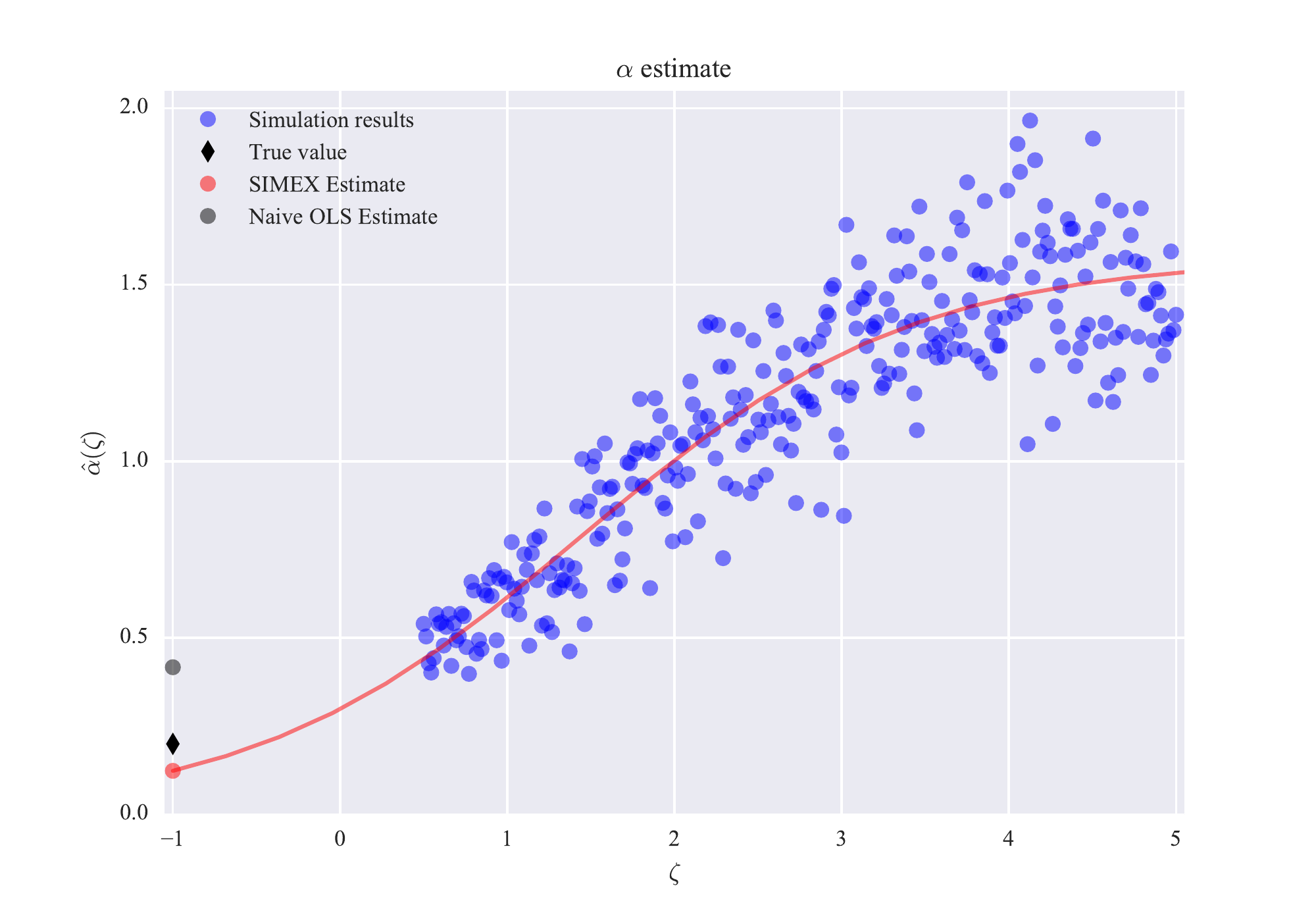}    % The printed column
\caption{Illustration of the SIMEX procedure of Section~\ref{SIMEX_section}. The error added to the measured data is indicated by the factor $\zeta$, with $\zeta=-1$ indicating the error-free state towards which simulation is extrapolated.  This figure illustrates one realisation of the simulations for $\alpha$.}  % width is 8.4 cm.
\label{fig:simex_illustration}                                % Size the figures
\end{center}                                 % accordingly.
\end{figure*}
\item Once the unbiased parameter estimates $\bm{\hat \theta}(\zeta=-1)$ were found by substitution into equations such as \equref{equ:alpha_logistic}, the errors relative to \tabref{tab:params} were calculated as

\begin{equation}
Error=\frac{\bm\theta-\bm{\hat\theta}(\zeta=-1)}{\bm\theta} \times 100.
\end{equation}
\end{enumerate}

\begin{table*}[tb]
\centering
\caption{Summary of distributional characteristics of parameter estimate errors for 300 random error realisations. These data are presented graphically in \figref{fig:results_params}.} 
\label{tab:results_params}
\begin{tabular}{c|ccc|ccc|ccc}
\hline 
\textbf{Method} &  & $\bm{\alpha}$ &  &  & $\bm{\phi_c}$ &  &  & $\bm{\epsilon}$ &  \\ 
 & \textbf{2.5\%} & \textbf{Mean} & \textbf{97.5\%} & \textbf{2.5\%} & \textbf{Mean} & \textbf{97.5\%} & \textbf{2.5\%} & \textbf{Mean} & \textbf{97.5\%} \\ 
\hline
Na\"ive & -188 & -91 & -9.21 & -245 & -162 & -58 & -459 & -286 & -123 \\ 

SIMEX & -23 & 39 & 73 & -108 & -16 & 57 & -173 & 54 & 26 \\ 
Bayes & -62 & -3 & 39 & -111 & -24 & 59 & -175 & -56 & 25 \\ 
\hline 
\end{tabular} 
\end{table*}

We recommend that calibration for M\&V purposes only be done using IEC-qualified meters. The calibration was simulated using the worst meter-CT combination that still conforms to an IEC specification (Class~3 meter and Class~5 CT) in order to be conservative. The overall accuracy of such a system, over the majority of the measurement range, is $\sqrt{0.03^2+0.05^2}=5.8\%$. One can see that the CT error dominates the overall uncertainty~\cite{birch2003measurement}. Replacing the meter in this system with a more accurate one will have little effect, reducing uncertainty to $5.4\%$ for a Class~2 meter. However, replacing the Class~5~CT with a Class~3~CT will reduce the overall uncertainty to approximately $4.24\%$.

Initially, we used LR on a smaller, approximately linear subset of the data, namely $\zeta \in [0,2]$. This worked well for $\alpha$ and $\epsilon$ estimates, but consistently overestimated $\phi_c$. The sigmoid shape was also partially hidden while we were using the discrete $\bm\zeta$ approach described in Step~\ref{second_regression} of Section~\ref{SIMEX_section}. If this approach is followed, the mean or mode of each $\zeta$ should be plotted rather than the full set, in order to show the shape of the data more clearly for regression model selection. However, we have found that a linearly spaced $\bm\zeta$ illustrates the shape of the function the best, as is seen in \figref{fig:simex_illustration}.

Selecting the right calibration period is important. If calibration is done over a weekend, for example, the proper power and power factor ranges will not be observed. Selecting a good calibration period is easy for a simulation study such as this one where all the data are available. However, it is more difficult in real situations when the data have not been observed yet. Therefore, the in-situ meter calibration period should be selected with care and in consultation with the facility manager. The IPMVP's recommendation for whole-building measurement, that ``all operating conditions be represented fairly'' during the baseline measurement period, should be followed. Furthermore, if Energy Conservation Measures (ECM's) are installed after the baseline period in an M\&V project, meter recalibration may be necessary, depending on the changes. The installation of Power Factor Correctors, which would decouple the power and power factor profiles, are an example of a case where baseline period parameter estimates may not hold during the reporting period.

\subsection{Discussion of Results} \label{discussion_of_results_params}

\begin{figure*}[tb]
\begin{center}
%\vspace{5cm}
\includegraphics[width=0.8\textwidth]{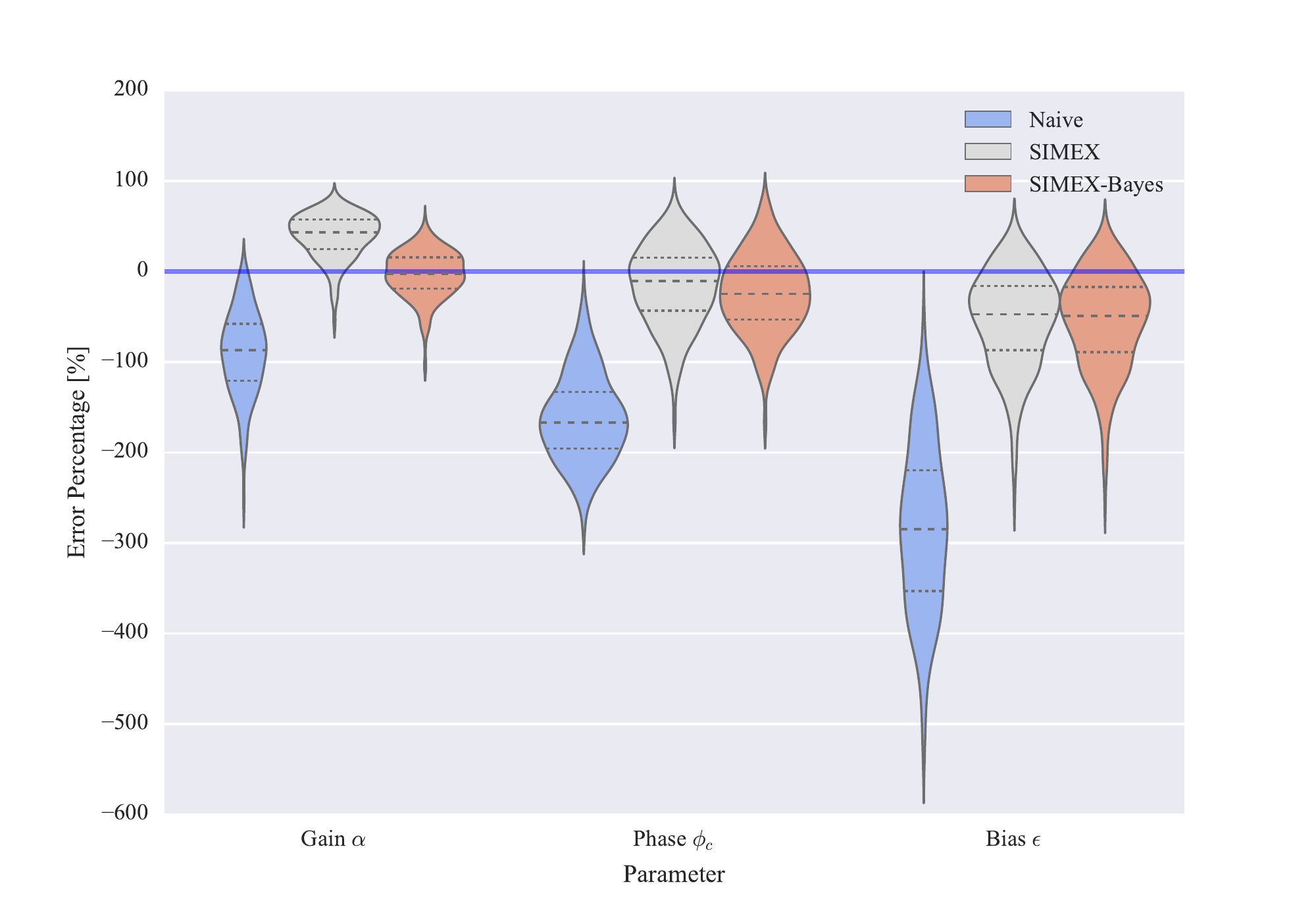}    % The printed column
\caption{Violin plot showing probability distribution shapes for Na\"ive, SIMEX, and SIMEX-Bayes parameter estimates, with quartiles indicated. A discussion of this figure can be found in Section~\ref{discussion_of_results_params}.}  % width is 8.4 cm.
\label{fig:results_params}                                 % Size the figures
\end{center}                                 % accordingly.
\end{figure*}

Although SIMEX is viable for this case, it does not un-bias parameter estimates perfectly: for certain realisations of random noise, such as where most points happen to be biased in the same direction, the starting data set for $\zeta=0$ is misleading, and SIMEX estimates will be imperfect. Therefore, to evaluate the reliability of the different methods, the process above was repeated for various realisations of $\mathbf{x}^*$ and $\mathbf{y}^*$ in~\equref{equ:x_star_given_x} and \equref{equ:carobbi_error_y_star}. Altogether 300 realisations were simulated, and a summary of the results are shown in~\tabref{tab:results_params} and in a violin plot in~\figref{fig:results_params}. This figure also shows the SIMEX-Bayes result for comparison. The SIMEX-Bayes method will be introduced and discussed in the next section.

A violin plot is similar to a box-plot in that it shows the probability distributions of the parameters. Where a box plot indicated the quartiles with a box and whiskers, a violin plot shows the full probability density function in mirrored form on a vertical axis.  The dashed line indicates the median, and the dotted lines the quartiles. Long, slender shapes such as for the Na\"ive bias estimate in \figref{fig:results_params} indicate a large variance and thus uncertainty in the estimate. Short, wide shapes like the SIMEX gain estimate indicate low variance and concentrated probability mass. Symmetric shapes such as for the SIMEX phase estimate indicate a symmetric probability distribution around the mean. Asymmetry such as for the SIMEX bias estimate indicates that the parameter estimates are skewed, in this case towards zero.

For \figref{fig:results_params}, estimates with zero (error) means will, on average, be error-free, although some variance is expected. This is the desirable result. The first notable observation is that the na\"ive estimates are further away from the zero line than the SIMEX estimates. This is to be expected: the na\"ive method is should be more biased, and this feature confirms the errors-in-variables theory. We also observe that the SIMEX estimate errors have smaller variances. This means that the SIMEX method converges on its less biased estimates more reliably. It is therefore more robust to the random effects of sampling than the ordinary least squares regression. The error in the $\epsilon$ estimate is the largest. However, to put it in perspective, a 100\% error  in $\epsilon$ means that $\hat \epsilon=10$ for $ \epsilon \sim Normal(5,2.5)$, given data in the range $(0,200)$. A 100\% error is therefore only a 2.5\% error relative to the data range. A 100\%~error in the gain $\alpha$ could be much more significant (representing a 100\% error relative to the data range), although a caveat to this assertion is discussed in Section~\ref{application}.

From these results we can see that the SIMEX procedure produces superior estimates to na\"ive regression, although they are not perfect. However, even if SIMEX produces better estimates on average, the quality of the prediction will depend on the specific combination of estimates in a specific set, and not only on the means across sets. A discussion of this result would be premature in this section, and the reader is referred to Point~\ref{heatmap_discussion} of Section~\ref{application}. In the next section, we will evaluate this interactive effect.

\section{Application to Meter Calibration} \label{application}

We will now compare the three meters used above based on how accurately they predict a longer measurement period than the calibration period. We consider three cases. The first is a laboratory-calibrated Class~3~meter with a Class~5~CT. This case is simply the readings of the reference instrument (calibrator) used for disciplining the other two meters. The second is a meter disciplined using the na\"ive procedure; assuming that the calibrator readings contain no error. The third is a meter disciplined using the SIMEX procedure, with Bayesian refinement. The parameter estimates obtained by disciplining the meter using the data from 2 February 2016 are then used to predict the energy consumption for the period 1 January 2016 - 3 August 2016.

Two goodness of fit metrics were selected to evaluate how well the predictions correspond to the true values for each of these 300 data sets. The Coefficient of Variation on the Root Mean Square Error (CV(RMSE)) and Normalised Mean Bias Error (NMBE) have been found to be the most popular criteria against which Calibrated Simulation M\&V model prediction goodness of fit is evaluated~\cite{granderson2016accuracy}. The NMBE measures whether the model consistently overpredicts or underpredicts energy use. The CV(RMSE) measures how closely the model tracks the actual data up and down: similar to its variance. An NMBE of 0\% would indicate no difference between the prediction and actual mean energy use, and a CV(RMSE) of 0\% would indicate no variance in the prediction relative to the actual.

For the calibrator, the CV(RMSE) happens to correspond to its combined precision of 5-6\%. However, the two metrics express uncertainty in slightly different ways and do not always correspond. Since we assume that the meter is unbiased, and specify it that way for the calibration, its NMBE is close to 0\%.

\begin{figure*}[t]
\begin{center}
%\vspace{5cm}
\includegraphics[width=0.8\textwidth]{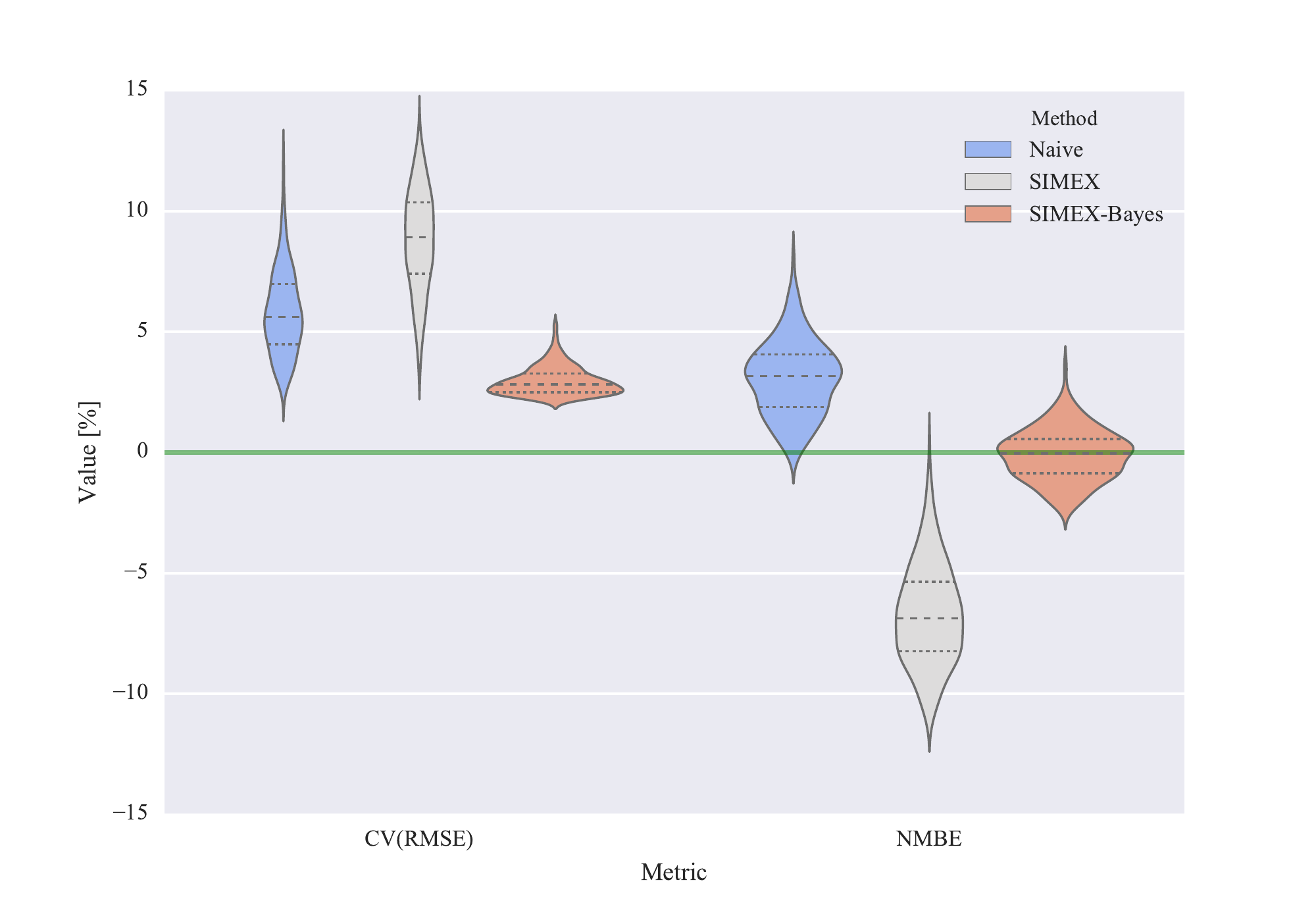}    % The printed column
\caption{Violin plot showing probability distribution shapes of goodness of fit metrics using parameter estimates of Na\"ive and SIMEX methods. Quartiles and median indicated by dashed lines. Two outliers were removed from the SIMEX plots to improve the vertical scale. A discussion of this figure can be found in Section~\ref{discussion_of_results_metrics}.}  % width is 8.4 cm.
\label{fig:results_metrics}                                % Size the figures
\end{center}                                 % accordingly.
\end{figure*}

This goodness of fit was evaluated in the following way: 
\begin{enumerate}
\item Generate observed energy use for the UUT ($\mathbf{y}^*$), for the full data set, by \equref{equ:carobbi_error_y_star}.
\item Generate observed energy use for the calibrator ($\mathbf{x}^*$), for the calibration period, using \equref{equ:x_star_given_x}.
\item Using only the 24-hour calibration data set, employ SIMEX and the na\"ive regression to estimate parameters $\alpha, \phi, $ and $\epsilon$. 
\item \label{heatmap_discussion} Refine SIMEX estimate through Bayesian regression. Although the parameter estimates of the SIMEX method are clearly superior to the na\"ive method, as shown in the previous section,~\figref{fig:results_metrics} shows that the resultant CV(RMSE) and NMBE on the rest of the data set are \textit{worse}. The reason is plotted in \figref{fig:CVRMSE_heatmap}. 
\begin{figure*}[tb]
\begin{center}
%\vspace{5cm}
\includegraphics[width=0.8\textwidth]{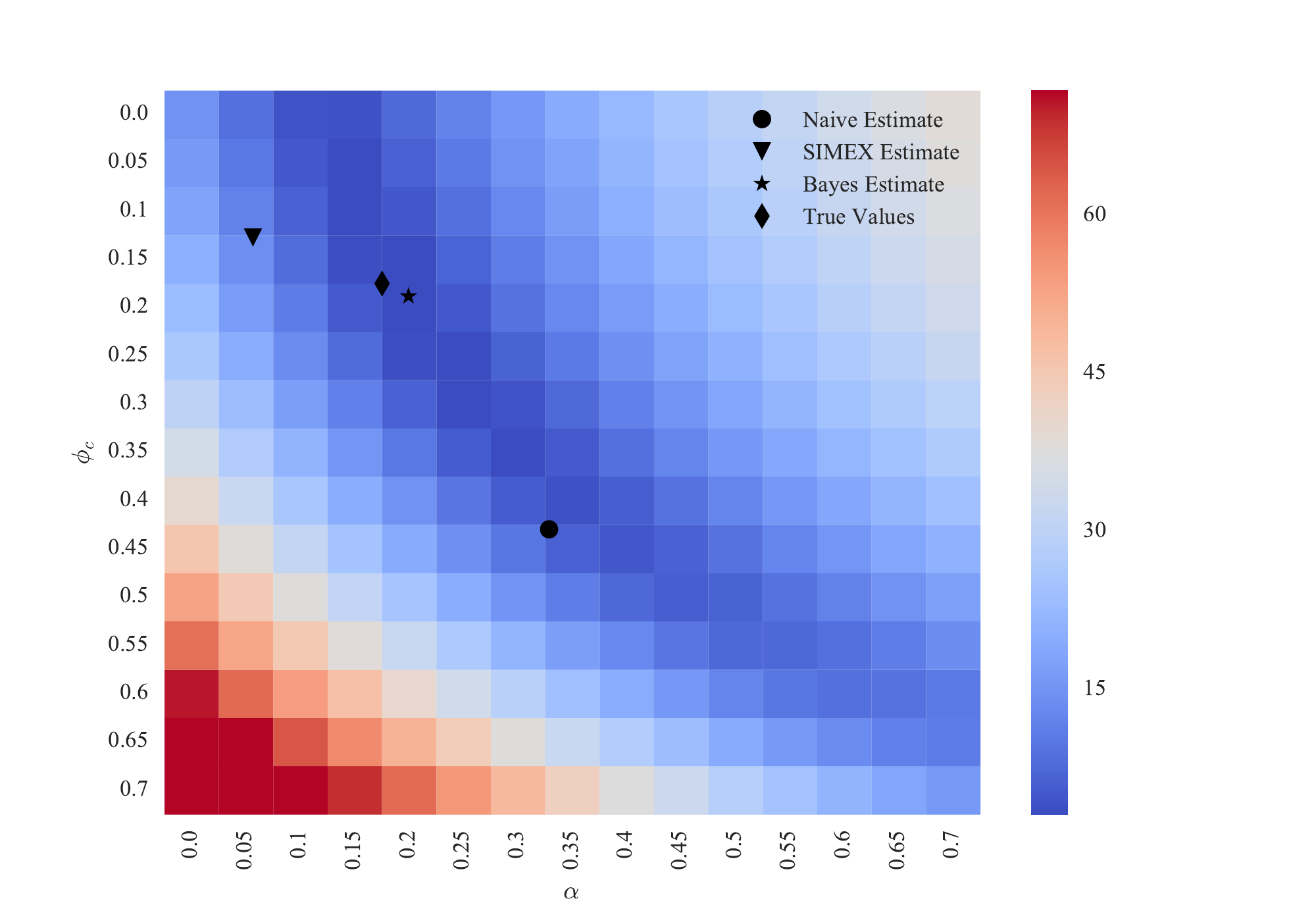}    % The printed column
\caption{CV(RMSE) (indicated by colour) for different combinations of parameters $\alpha$ and $\phi_c$. The parameter combinations plotted are for single instances of solutions. This plot assumes a bias error $\epsilon=5$. The positions of the SIMEX and Bayes estimates relative to the true values varies from realisation to realisation. A discussion of this figure can be found in Section~\ref{discussion_of_results_metrics}.}  % width is 8.4 cm.
\label{fig:CVRMSE_heatmap}                                % Size the figures
\end{center}                                 % accordingly.
\end{figure*}
Although the na\"ive estimates of the parameters are much worse than the SIMEX estimates, the prediction quality (goodness of fit) is dependent on their combination. Thus $\alpha$ may be overestimated and $\phi_c$ underestimated, but they cancel each other out in such a way that the final result is close to the true value, especially with noise in $\epsilon$ adding some tolerance to the results. Neglecting $\epsilon$ for a moment, we can visualise this as in \figref{fig:CVRMSE_heatmap}. Gain error is the \textit{x}-coordinate on the map, phase error is the~\textit{y}-coordinate, and CV(RMSE) is the height, indicated by colour. Low CV(RMSE) values form a low CV(RMSE) valley running northwest to southeast. Altough there is only one coordinate that is ``correct" in the sense of corresponding to the true values, this valley indicates the combinations of gain and phase error values that will also yield a low CV(RMSE). Now, because the sum-squared error is a major component of the CV(RMSE) calculation, a low sum squared error will lead to a low CV(RMSE). Least Squares regression finds a solution with the least sum of squares error. In other words, the na\"ive method effectively optimizes for CV(RMSE), and we are therefore not surprised that it produces results with low CV(RMSE)'s, even if the individual parameter values themselves are not accurate. This lack of convergence on the true values shows a parameter identifiability problem between the gain and phase errors $\alpha$ and $\phi_c$ in \equref{equ:carobbi_error_y_star}. Another confounding factor is that the power factor $\phi$ is correlated with energy use as referred to earlier. This correlation, as well as the small range for $\phi$, do not help identifiability.

Because the SIMEX method improves the parameter estimates independently, it does so without considering their combined effect on the sum squared error of the fit. This results in more accurate estimates but slightly higher CV(RMSE) values. We therefore chose to refine SIMEX estimates using Bayesian regression. This step changes the SIMEX estimates slightly to serve the double purpose of improving the goodness of fit metrics and providing probability distributions on the parameter estimates. These distributions can be used for risk and uncertainty quantification calculations, both on the parameter estimates and also on the prediction energy use. As shown in \figref{fig:CVRMSE_heatmap}, the Bayesian method does not necessarily interpolate linearly between the SIMEX estimates and true values. However, it does converge on parameter estimates in the SIMEX region while yielding improved CV(RMSE) and NMBE values. The method is explained more fully in Section~\ref{section_Bayes}. Using the Bayesian method on the na\"ive estimates, or using the na\"ive optimisation algorithm with the SIMEX estimates as its starting position, did not improve on the original na\"ive estimates.

\item \label{prediction_step} Generate predicted energy use for full data set by inverting \equref{equ:carobbi_error_y_star} using the parameter estimates, so that:
\begin{equation}
\mathbf{x}_{predicted}=\frac{\mathbf{y}^*-\hat\epsilon}{(1+\hat\alpha)cos(\phi+\hat\phi_c)}
\end{equation}

\item As with the calibration procedure in Section~\ref{SIMEX_application}, repeat Steps~1-5 300 times to account for different random realisations of $\mathbf{x}^*$ and $\mathbf{y}^*$. The summary statistics of the goodness of fit metrics from these simulations are given in \tabref{tab:results_metrics}, and plotted in \figref{fig:results_metrics}.
\end{enumerate}

Before the results are discussed, an explanation of the Bayesian refinement is given.

%The first meter will be called the reference meter. The precision specification on this meter will be at the~95\% confidence level, specified at the base current level~$I_{N} = 200A$. For Class~5 with we therefore have an accuracy of 5A, which is 1.15~kWh at~230V for an inductive power factor of 0.5 to 1 in the range 0.2$I_{N}$ to $I_{MAX}$. We assume that the CT accuracy is uncorrelated to the meter accuracy, and thus their combined accuracy is their RMS, not their sum.

\begin{table*}[tb]
\centering
\caption{Summary of distributional characteristics of two goodness of fit metrics for the methods under investigation: the Coefficient of Variation on the Root Mean Square Error (CV(RMSE)), and the Normalised Mean Bias Error (NMBE). These results are presented graphically in \figref{fig:results_metrics}.}
\label{tab:results_metrics}
\begin{tabular}{c|ccc|ccc}
\hline 
\textbf{Method} &  & \textbf{CV(RMSE)} &  & &\textbf{NMBE} &  \\ 
 & \textbf{2.5\%} & \textbf{Mean} & \textbf{97.5\%} & \textbf{2.5\%} & \textbf{Mean} & \textbf{97.5\%}\\ 
\hline
Na\"ive & 3.03 & 5.8 & 9.91 & 0.33 & 3.08 & 6.34\\ 

SIMEX & 4.59 & 8.87 & 12.49 & -10.344 & -6.79 & -2.33\\ 

Bayes & 2.27 & 2.96 & 4.35 & -2.05 & -0.09 & 2.03\\ 

\hline 
\end{tabular} 
\end{table*}

\subsection{Bayesian Refinement}\label{section_Bayes}

Bayesianism is a branch of statistics in which conditional probabilities are derived from distribution theory by the laws of logic. A full exploration of Bayesian theory is beyond the scope of this paper, and the reader is referred to Gelman et al.~\cite{gelman2014bayesian} and Kruschke~\cite{kruschke2015doing} for more detailed information. Briefly, it is named after Bayes theorem, which states that the posterior probability of the parameter values $\bm{\theta}$ given the data observed $\mathbf{D}$, can be expressed in terms of known probabilities. These are the likelihood of the data given some parameter function $\pi(\mathbf{D}|\bm\theta)$, and a `prior' probability for the parameter values $\pi(\bm{\theta})$. Mathematically, it is expressed as:

\begin{equation} \label{equ:Bayes}
 \pi(\bm{\theta}|\mathbf{D}) = \frac{\pi(\mathbf{D}|\bm\theta)\pi(\bm{\theta})}{\pi(\mathbf{D})}.
 \end{equation} 
 
The increase in computing power and the derivation of useful numerical techniques such as Markov Chain Monte Carlo (MCMC) has solved two of the great difficulties in Bayesian analysis. These are the intractability of analytical solutions to non-trivial problems, and the difficulty in specifying the $\pi(\mathbf{D})$ term. Because of MCMC, the application of Bayesian theory has developed into a powerful, intuitive statistical and machine learning tool.

In a Bayesian framework,  all model parameters are treated as unknown random variables, and the data are regarded as realisations of these distributions. 

The modes of the posterior distributions for $\alpha,\phi_c,$ and $\epsilon$ will correspond to their maximum likelihood estimates given the data observed. We observe
\begin{equation} \label{equ:y_star_given_x_star_theta}
\pi(\mathbf{D}|\bm\theta) = \pi(\mathbf{x}^{*}, \mathbf{y}^{*} \mid\alpha, \phi_c, \epsilon, I)
\end{equation}
where $I$ is the prior information at our disposal through the SIMEX result, and $\alpha,\phi_c,$ and $\epsilon$ are unknown. By Bayes' theorem in~\equref{equ:Bayes}, through a numerical algorithm, this can be inverted so that the posterior conditional probability estimates of the parameters
\begin{equation}
\pi(\bm\theta | \mathbf{D})=\pi(\alpha, \phi_c,\epsilon \mid \mathbf{x}^{*}, \mathbf{y}^{*}, I)
\end{equation}
are found.

\subsubsection{Prior Selection}
Before a Bayesian model can be solved, the priors have to be specified. To let the model be as objective as possible, priors are often specified to be vague or non-informative. The specification of priors can be contentious regardless of what is selected. Overconfident priors can bias the posterior distributions, especially for cases where few data points are available. Non-informative priors do not bias the posterior distribution (or bias it towards the data). However, this approach has also drawn criticism as non-informative posteriors elicited in this way can be unhelpful~\cite{gelman2008weakly, berger2006case}. In energy studies, informative priors based on previous studies have often been used and enjoy a strong precedent~\cite{heo2011bayesian, lee2013towards, booth2013decision, heo2015evaluation}. In the empirical Bayesian approach, priors can be informed by prudent use of the data itself, such as  $\hat{\bm\theta}_{SIMEX}$ obtained from the SIMEX algorithm. However, care must be taken when selecting data-dependent priors, as these can lead to a case of ``data reinforcing data". This results in misleadingly high confidence on posterior estimates. Nevertheless, when such techniques are used correctly, they do have precedent~\cite{gelman2008weakly}, and are mathematically defensible in certain cases. This has been shown by Darnieder in his PhD thesis on the topic~\cite{darnieder2011bayesian}. In our case, specifying vaguely informative priors is justified because the SIMEX parameter estimates do not arise naturally from the data itself, the way it would when using the mean as a prior in a model that estimates the mean. We use the priors to `constrain' the algorithm to the solution space around the SIMEX solution. If overly vague priors are specified, the algorithm tends to converge on low CV(RMSE) solutions far away from the SIMEX estimates, and thus far away true values.The priors on the parameters are specified as follows:

\begin{equation}
\pi(\alpha) \sim Normal(\mu=\hat\alpha_{SIMEX},\ \sigma = 5),
\end{equation}

\begin{equation}
\pi(\phi_c) \sim Normal(\mu=\hat\phi_{c, SIMEX},\ \sigma = 1),
\end{equation}

\begin{equation}
\pi(\epsilon) \sim Normal(\mu=\hat\epsilon_{SIMEX},\ \sigma = 5).
\end{equation}

We also specify a prior on $\mathbf{x}^*$. If the meter errors were Berkson, this prior would be perfectly representative. However, since the error is located in the meter itself, they are classical. Therefore the prior below is not perfect but does allow for variation in $\mathbf{x}^*$ so that the model does not consider the observed values for $\mathbf{x}^*$ as fixed. The prior on $\mathbf{x}^*$ is specified as

\begin{equation}
\pi(\mathbf{x}^*) \sim Normal(\mu=\mathbf{x}^*,\ \sigma = \sigma_u).
\end{equation}

We define the likelihood function $\pi(\mathbf{D}|\bm\theta)$ as a multivariate Student-T~distribution. The thicker tails of this distribution allows for more robust inference, since outliers have a smaller effect on the posterior mean~\cite{Lange1989robust}. In this case, our data are the values observed from the reference and the UUT meters, and our priors are the SIMEX parameter estimates. Therefore:
%\begin{equation}
%\pi(\hat\alpha(\bm\zeta)|\alpha)\sim StudentT(\hat\alpha(\bm\zeta) | %\mu=a_\alpha\bm\zeta+b_\alpha,\ \sigma=\sigma_p,\ \nu=\nu_p),
%\end{equation}

\begin{equation}
\pi(\mathbf{y}^*|\mathbf{x})\sim StudentT(\mathbf{y}^* | \bm\mu= \bm\mu_p, \bm\sigma=\pi(\sigma_p),\ \bm\nu=\pi(\nu_p))
\end{equation}
%\begin{eqnarray}
%\begin{array}{l l}
%\pi(\mathbf{y}^*|\mathbf{x})\sim StudentT(\mathbf{y}^* |
%&\bm\mu= \bm\mu_p,\\ &\bm\sigma=\pi(\sigma_p),\\
%&\bm\nu=\pi(\nu_p))
%\end{array}
%\end{eqnarray}
where 
\begin{equation}
\bm\mu_p = (1+\pi(\alpha))\pi(\mathbf{x}^*)cos(\phi+\pi(\phi_{c}))+\pi(\epsilon),
\end{equation}
as in \equref{equ:carobbi_error_y_star} and the hyperpriors are defined as
\begin{equation}
\pi(\nu_p) \sim Exponential(48^{-1})
\end{equation}
and
\begin{equation}
\pi(\sigma_p) \sim HalfCauchy(1).
\end{equation}

The choice of `48' as the inverse scale parameter for the exponential distribution relates the the number of data points in the calibration period~\cite{kruschke2015doing}. For the scale parameter $\sigma$, we follow Gelman's recommendation of a half-Cauchy distribution~\cite{gelman2006prior}.
\subsubsection{Solving the model}
Although a full Bayes-MCMC is standard, Automatic Differentiation Variational Inference (ADVI)~\cite{kucukelbir2016automatic} is a new and much faster alternative to standard MCMC algorithms. It has comparable accuracy and is useful for batch runs where the different approaches are compared for different error realisations on the same data set. The model is solved using 50,000 runs of the ADVI algorithm. The starting points are specified as as the SIMEX estimates. The analysis is performed in Python via the \textprogfont{PyMC3}~\cite{salvatier2016probabilistic} library. Because only point estimates of the parameters are of interest for the current problem, we did not utilise the full Bayesian capability of eliciting full posterior probability distributions for each of the runs.

\subsection{Discussion} \label{discussion_of_results_metrics}

The resultant CV(RMSE) and NMBE for the Na\"ive and SIMEX calibrated meters are shown in \tabref{tab:results_metrics} and \figref{fig:results_metrics}. In these, it can be seen that the Bayesian refinement improves the CV(RMSE) SIMEX estimates substantially, from 8.87 to 2.96. The average NMBE improves from -6.79\% to -0.09\%. A CV(RMSE) of 2.96\% seems lower than the original 5.8\% noise in the data. However, one should bear in mind that although CVRMSE is the appropriate metric to use, it cannot be compared to the way in which the noise is expressed originally. From Equation~4-4 of ASHRAE~Guideline~14~2014~\cite{Ashrae2014} for $\alpha$,

\begin{equation}
\textrm{CV(RMSE)}_{\alpha} = \frac{\sqrt{\frac{\sum (\alpha_i-\hat \alpha)^2}{n-par}}}{\bar \alpha}
\end{equation}

where $y_i$ is the true value, $\hat y_i$ is the model estimate, $\bar y$ is the mean, $n$ is the number of data points, and $par$ is the number of parameters. As the name suggests, it is therefore the mean of the sum squared error, normalised with respect to the mean of the data. This is a different value to the relative precision of the meter.

\figref{fig:results_metrics} shows that the Bayes-SIMEX procedure produces predictions with superior goodness of fit, both in terms of bias and in terms of CV(RMSE). Besides the violin plot, it is also graphically illustrated in~\figref{fig:CVRMSE_heatmap}, where the SIMEX-Bayes coordinate approaches the true coordinate. The distributions are also tighter than for the other procedures, indicating improved consistency compared to SIMEX and na\"ive regression. \figref{fig:results_params} indicates that Bayes-SIMEX does not do this at the cost of individual parameter estimates. On the contrary, superior and more consistent parameter estimates are also obtained.

To put these values in perspective, the ASHRAE Guideline 14-2014 requires an NMBE below 5\% for monthly data and 10\% for hourly data~\cite{Ashrae2014}. CV(RMSE) requirements are 15\% and 30\% respectively. As this is half-hourly data, the requirements are in effect even more generous. However, it should be kept in mind that the ASHRAE metrics do not refer to the calibration of measured energy data, but to building energy modelling requirements \textit{relative~to} measured energy data. The calibration figures in this paper are therefore baselines to which traditional M\&V modelling uncertainty is added, before being compared to ASHRAE requirements. Nevertheless, the calibration procedure is so effective, even with low accuracy meters and only 24 hours of calibration, that building models on energy use data obtained from this calibration method should still be acceptable. With longer calibration times or more accurate calibrators, these figures could also improve.

We should note that valid calibration requires more than simply having a reference instrument available. An adequate quality system needs to be followed to ensure that results are traceable and repeatable. However, we may conclude that from a technical point of view, the calibration itself does not require exceptionally accurate instruments for practical~M\&V purposes, and can reduce monitoring costs significantly through in-situ calibration.

\section{Conclusion}

The calibration of energy meters for monitoring projects can be expensive, and may not be cost-effective in terms of the gains in accuracy. We propose disciplining or verifying an uncalibrated meter in-situ by using another calibrated commercial-grade metering system, in this case, a Class~3 meter and a Class~5 Current Transformer (CT). By using the Simulation Extrapolation Measurement Error Model and refining parameter estimates using a Bayesian approach, the verified meter is shown to report energy use accurately and with low error variance compared to na\"ive Ordinary Least Squares methods. For the data set under investigation, the Coefficient of Variation on the Root Mean Squared Error was reduced from 8.87\% to 2.96\%, and the Normalised Mean Bias Error from -6.79\% to -0.09\%. To be conservative, the most inaccurate meter-CT combination for IEC-qualified instruments was selected, and has been demonstrated to have acceptable accuracy. For any other combination of IEC-qualified meters and CTs, more accurate results should be obtained if calibration period data is representative. The general method proposed in this paper may also be applied to instruments other than energy meters.

\section*{Acknowledgement}
We wish to thank our anonymous reviewers and the editor for their helpful comments.
The financial assistance of the National Research Foundation (NRF) and the National Hub for the Postgraduate Programme in Energy Efficiency and Demand Side Management towards this research is hereby acknowledged. Opinions expressed and conclusions arrived at are those of the authors, and are not necessarily to be attributed to the NRF.

%In most electricity meter sampling cases where sampling CV is not very low, the cost of higher accuracy (e.g.~Class~0.2S) metering is not justified by the increased reporting accuracy. A larger sample of the lower accuracy electricity meters will be of more benefit to minimise uncertainty. However, for projects using measurement systems other than electricity meters, the contribution of measurement uncertainty should be calculated, as it may make a material difference to the overall uncertainty.
%For Energy~Performance~Contracting projects using the guaranteed savings model, it is found that more accurate meters do not necessarily increase the project's Expected~Value for the project developer or the owner, and does not contribute meaningfully to the determination of a guaranteed savings figure. A Bayesian method for the calibration of an energy meter has also been found to be more accurate than standard frequentist approaches, yielding more accurate and complete parameter estimates.

%\begin{thebibliography}
\bibliographystyle{elsarticle-num} 
\bibliography{lib}






%\bibitem{}

%\end{thebibliography}
\end{document}